\documentclass[twocolumn,showpacs,aps,prl,superscriptaddress]{revtex4}

\usepackage{graphicx}
\usepackage{dcolumn}
\usepackage{amsmath}
\usepackage{epsfig}

\input pubboard/babarsym

\newcommand{\BABARPubYear}    {06}
\newcommand{\BABARPubNumber}  {035}

\newcommand{\SLACPubNumber} {11923}

\def\figurebox#1#2#3{%
    \def\arg{#3}%
    \ifx\arg\empty
    {\hfill\vbox{\hsize#2\hrule\hbox to #2{\vrule\hfill\vbox to #1{\hsize#2\vfill}\vrule}\hrule}\hfill}%
    \else
    {\hfill\epsfbox{#3}\hfill}%
    \fi}

\def\Resultphirhozeroformal{5.7\pm 0.5(\mathrm{stat}) \pm 0.8(\mathrm{syst}) ~\mathrm{fb}}
\def\Resultrhozerorhozeroformal{20.7 \pm 0.7 (\mathrm{stat})\pm 2.7(\mathrm{syst}) ~\mathrm{fb}}

\begin{document}
\preprint{\babar-PUB-\BABARPubYear/\BABARPubNumber} 
\preprint{SLAC-PUB-\SLACPubNumber} 

\begin{flushleft}
\babar-PUB-\BABARPubYear/\BABARPubNumber\\
SLAC-PUB-\SLACPubNumber\\
\vspace{1cm}
\end{flushleft}

\title
{
{\large \bf 
Observation of $\epem$ Annihilations into the $\boldmath{C=+1}$ Hadronic Final States  $\rho^0\rho^0$ and $\phi\rho^0$  
} 	
}

%
\author{B.~Aubert}
\author{R.~Barate}
\author{M.~Bona}
\author{D.~Boutigny}
\author{F.~Couderc}
\author{Y.~Karyotakis}
\author{J.~P.~Lees}
\author{V.~Poireau}
\author{V.~Tisserand}
\author{A.~Zghiche}
\affiliation{Laboratoire de Physique des Particules, F-74941 Annecy-le-Vieux, France }
% ...
\collaboration{The \babar\ Collaboration}
\noaffiliation

\date{\today}

\begin{abstract}

We report the first observation of \epem annihilations into
states of positive $C$-parity, 
namely $\rho^0 \rho^0$  and $\phi\rho^0$. 
The two states 
are observed in the $\pip\pim\pip\pim$ and $K^+K^-\pip\pim$
final states, respectively,
in a data sample of 225~\invfb  collected by the \babar~experiment at 
the  \pep2 $e^+e^-$ storage rings at energies near 
$\sqrt{s}\! ~=$ 10.58~\gev. 
The distributions of  $\cos\theta^*$, where $\theta^*$ is the center-of-mass polar angle of the $\phi$ meson  or 
the forward $\rho^0$ meson, suggest production  by  two-virtual-photon annihilation. 
We measure cross sections within the range 
$|\cos\theta^*|\! ~<~ \! 0.8$  of 
$\sigma(\epem\!\! \to\! \rho^0 \rho^0)\! ~=~$$\Resultrhozerorhozeroformal$ and
$\sigma(\epem\!\! \to\! \phi \rho^0)\! ~=~$$\Resultphirhozeroformal$.

\end{abstract}


\maketitle

\begin{figure}
\begin{center}
\includegraphics[width=6.7cm]{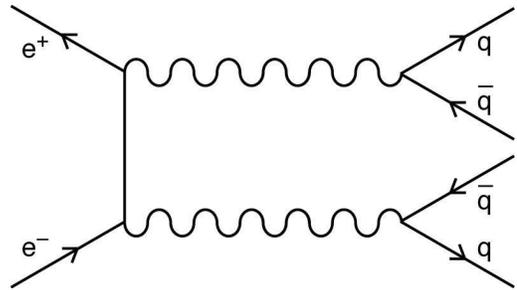}
\caption{\small Two-virtual-photon annihilation diagram. }
\label{fig:twovirtualphoton}
\end{center}
\end{figure}

The process $\epem\rightarrow \mathrm{hadrons}$ at center-of-mass (CM) energy $\sqrt{s}$  
far below the $Z^0$ mass is dominated by annihilation via a single virtual photon 
with charge-conjugation parity $C=-1$. 
The high luminosity of the  $B$ factories provides  an opportunity to explore 
rare, low  multiplicity final states with $C=+1$ 
such as those produced
in the two-virtual-photon annihilation (TVPA) process
depicted in Fig.~\ref{fig:twovirtualphoton}.  
The TVPA process has been ignored in the interpretation 
of the total hadronic cross section in $e^+e^-$ annihilations as input to 
calculations~\cite{g-2} of the muon $g\! -$2  and the running QED coupling $\alpha$. 
We report the first observation of the exclusive reactions 
$\epem \!\!\to\! \rho^0\rho^0$ and $\epem \!\!\to\! \phi\rho^0$,  
in which the final states are even under charge conjugation,
and therefore cannot be produced via single-photon annihilation.

This analysis uses a 205 fb$^{-1}$ data sample of \epem collisions collected  
on the \Y4S resonance and 20 
fb$^{-1}$ collected 40\mev below 
with the \babar~ 
detector at the SLAC PEP-II asymmetric-energy $B$ factory. 
The \babar~ detector is described in detail elsewhere~\cite{babardetector}.
Charged-particle momenta and energy loss are measured in the 
tracking system which consists of   
a silicon vertex tracker (SVT) and a 
drift chamber (DCH). 
Electrons and photons  are detected in a CsI(Tl)  calorimeter (EMC). 
An internally reflecting ring-imaging Cherenkov detector (DIRC) 
provides charged particle identification (PID). 
An instrumented magnetic flux return (IFR) provides identification of muons.
Kaon and pion candidates are identified using likelihoods of   
particle hypotheses calculated from the specific ionization in the DCH and SVT, 
and the Cherenkov angle measured in the DIRC. Electrons are  identified 
by the ratio of the  energy deposited in 
the EMC to the  momentum and by the shower shape; 
muons are  identified by the depth of penetration into the IFR.

\begin{figure}
\begin{center}
\includegraphics[width=8.8cm]{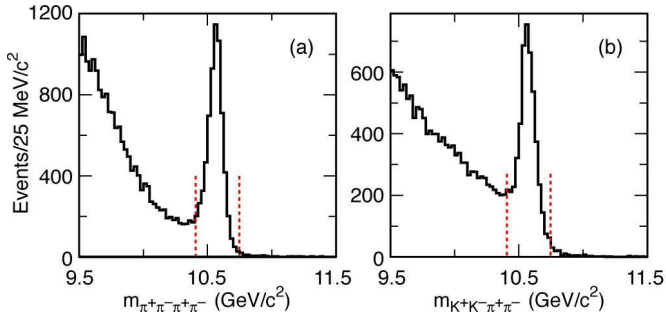}
\caption{\small 
Distributions of the invariant mass ($\Upsilon(4S)$ data) for the 
 a) $\pip\pim\pip\pim$ and b) $K^+K^-\pip\pim$ final states.
The accepted signal regions are indicated  by the dashed lines.
}
\label{fig:4trkmass}
\end{center}
\end{figure}

Events with four well-reconstructed charged tracks   
and a  total charge of zero are selected. 
Charged tracks are required to have at least 12 DCH hits and   
a polar angle in the range   
$0.41<\theta<2.54$ radians. The momenta of kaon and pion candidates are required to be  greater than 
800 and 600 \mevc, respectively. 
Among the four selected tracks, two oppositely charged tracks  must   be identified
as pions, and the other pair must  be identified as two pions or two kaons.
Events in which one or more pion candidates  
are identified as an electron or muon are rejected (lepton veto). 
We fit the four tracks to a common vertex, and require the $\chi^2$
probability to exceed 0.1\%. 
We accept events with reconstructed invariant mass   within 170~\mevcc of the
nominal  CM energy (Fig.~\ref{fig:4trkmass}).

\begin{figure}
\begin{center}
\includegraphics[width=8.8cm]{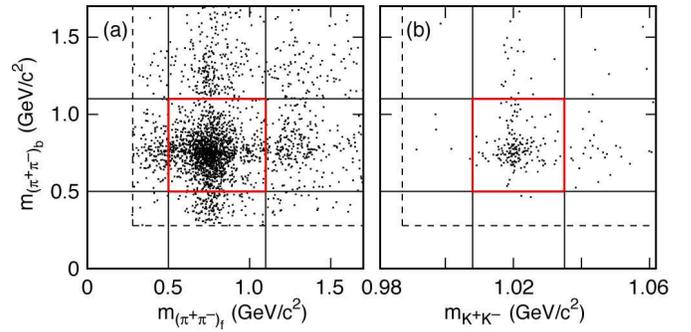}
\caption{\small 
Scatter plots of the invariant masses of the two oppositely charged
 pairs in the a) $\pip\pim\pip\pim$ and b) $K^+K^-\pip\pim$ final states. 
The dashed lines indicate $K^+K^-$/$\pi^+\pi^-$ thresholds. 
The solid lines show the nine tiles used in the fit.
}
\label{fig:myscat}
\end{center}
\end{figure}

 In the process  $e^+e^-\rightarrow \pi^+\pi^-\pi^+\pi^-$,
there are two possible pairings of $\pi^+$ mesons with $\pi^-$ mesons.  
However, only one combination 
appears in the  kinematic region of interest ($m_{\pi^+\pi^-}<2~\gevcc$) for both pairs.  
We label the pion pair 
with CM momentum vector pointing into the hemisphere defined by  the $e^-$ beam 
direction  $\pi^+\pi^-_{f}$ and  
the other as $\pi^+\pi^-_{b}$. 
Figure~\ref{fig:myscat}(a) shows the scatter plot
                 of the invariant masses of $\pi^+\pi^-_f$ and $\pi^+\pi^-_b$ 
from \epem\to$\pi^+\pi^-\pi^+\pi^-$ events, and
                 Fig.~\ref{fig:myscat}(b) the plot of invariant masses of $K^+K^-$ and $\pi^+\pi^-$
                 pairs from \epem\to$K^+K^-\pi^+\pi^-$ events. We observe correlations
                 of masses in Fig.~\ref{fig:myscat}(a) indicating the production of $\rho^0\rho^0$ 
                 final states, and in Fig.~\ref{fig:myscat}(b) indicating the production
                 of $\phi\rho^0$ final states.

To extract the number of $\epem\rightarrow \rho^0\rho^0$ and $\phi\rho^0$ signal events,
we perform a binned maximum-likelihood fit for  nine rectangular regions (tiles) in the two-dimensional
mass distributions, as shown in Fig.~\ref{fig:myscat}.
The signal box is the central tile (tile 5), defined by the mass ranges  
$0.5<m_{\pi^+\pi^-}<1.1$ GeV/c$^2$ and $1.008<m_{K^+K^-}<1.035$ GeV/c$^2$. 
For \epem\to$K^+K^-\pip\pim$, the expected number of events, $n_i$, for each tile $i$
can be expressed as:
\begin{equation}
n_i=f_i^S{S}+f_i^{\phi }N_{\phi }+f_i^{\rho^0 }N_{\rho^0 }+f_i^B{B},
\end{equation}
where $S$ is the  number of $\phi\rho^0$ signal events,
$N_{\phi }$ is the number of $\phi X$ background events,
$N_{\rho^0 }$ is the number of $\rho^0 X$ background events, 
and $B$ is the number of residual background events, in all nine tiles. 
The parameter $f_i^T$ is the fraction of events of type $T$ that contributes to tile $i$.
The signal fractions $f_i^S$ are modeled by Monte Carlo (MC) simulation~\cite{MC},  
and $f_i^{\phi }$ and $f_i^{\rho^0 }$ are obtained from 
the $\phi X$ and $\rho^0 X$ background shapes, which are estimated by fitting the
projections of $m_{K^+K^-}$ and $m_{\pi^+\pi^-}$ as described later. 
The residual background fractions $f_i^B$ are modeled by a linear function that can be expressed as  
\begin{equation}
f^B_i=\frac{\Delta x_i \Delta y_i [1+s_{\rho^0}(x_i-x_5) + s_{\phi}(y_i-y_5) ]}{\sum_{j=1}^{9} \Delta x_j \Delta y_j },  
\end{equation}
where $\Delta x_i $ and $\Delta y_i$ are the 
kinematically accessible 
dimensions of tile $i$, 
$x_i$ and $y_i$ are at the center of tile $i$, and $s_{\rho^0}$ and $s_{\phi}$ are  slopes 
obtained from the fits. 
A similar expression is used for the $\pip\pim\pip\pim$ case, where   
$\phi$ and $\rho^0$ are replaced with $\rho^0_f$ and $\rho^0_b$.

\begin{figure}
\includegraphics[width=8.8cm]{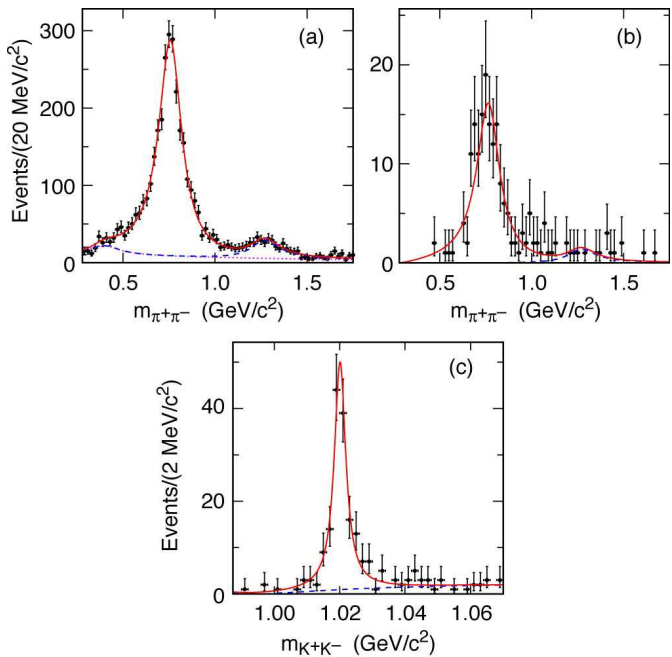}
\caption{\label{fig:threeplot} 
  Mass distribution for 
  a) $\pi^+\pi^-$ pairs in $\rho^0 \pi^+\pi^-$ events,
  b) $\pi^+\pi^-$ pairs in $\phi \pi^+\pi^-$ events, and
  c) $K^+K^-$ pairs  in   $\rho^0 K^+K^-$ events. 
The solid curves are 
the total fit. For the $\pi^+\pi^-$ cases, the dotted curve is the $\mu^+\mu^-$ component, 
while the sum of $f_2(1270)$ and $\mu^+\mu^-$ contributions are shown as dashed. 
For the $K^+K^-$ case, the dashed curve represents the threshold function. 
}
\end{figure}

The background fractions are obtained by mass projection fits which are confined 
to the central horizontal 
or central vertical $\phi$ or $\rho^0$ resonance band. 
The effect of neglecting the resonance width outside the central band, 
checked by smearing the background fractions in the central band 
into the adjacent tiles using the resonance widths 
obtained from MC, is found to be negligible. 
The mass projections in the central bands for $\pi^+\pi^-$ recoiling against a 
selected $\rho^0$ or 
$\phi$ 
and for $K^+K^-$ recoiling against a $\rho^0$ are 
shown in Fig.~\ref{fig:threeplot}. 
For  the $\rho^0\pi^+\pi^-$ case we fit the $\pi^+\pi^-$ mass projection to the sum of a 
$\rho^0$ component, an $f_2(1270)$ component, and a $\mu^+\mu^-$ background 
component. The $\rho^0$  is represented by the product of a P-wave relativistic Breit-Wigner
with its width set to  the Particle Data Group (PDG)~\cite{pdg04} value, a 
phase space term, and a factor  $1/m^2_{\pi\pi}$ due to production via a 
virtual photon. The $f_2(1270)$ is represented by a D-wave relativistic Breit-Wigner
with its  mean and width set to  the PDG values. The $\mu^+\mu^-$ background
shape is obtained from a sample of the  related channel $e^+e^-\rightarrow\rho^0\mu^+\mu^-$ 
isolated by requiring two  oppositely  charged tracks identified as muons.
For the $\phi\pi^+\pi^-$ case, 
we use the same background parameterization in terms  of $f_2$ and $\mu^+\mu^-$, 
but refit for their normalizations. For the $\rho^0 K^+K^-$ case, 
we fit the $K^+K^-$ mass  
projection to the sum of a Breit-Wigner with 
mean and width fixed to their PDG values for the $\phi$ signal, 
and a threshold function $(q^3)/(1+q^3R)$, where $q$ is kaon momentum in the $\phi$ 
rest-frame and $R$ is a shape parameter, for background.  
Assuming the masses of the two pairs to be uncorrelated 
and excluding the $\rho^0$ and $\phi$ signal contributions, the 
fitted functions are integrated to obtain  the tile 
fractions $f_i^{\rho^0}$, $f_i^{\phi}$, and $f_i^{\rho^0}$.

The extracted $\rho^0\rho^0$ and $\phi\rho^0$ yields in the signal box  are 
$1243\pm 43$ and $147\pm 13$ events, to be compared with  total of 
1508 $\pi^+\pi^-\pi^+\pi^-$ ($\sim18\%$ background)
and 163 $K^+K^-\pi^+\pi^-$ ($\sim10\%$ background) events in 
the signal box, respectively.

To investigate the possibility of $\rho^0\rho^0$ and  $\phi\rho^0$  
production in $\Upsilon(4S)$ decay, 
we examine the data recorded at  and 
below the $\Upsilon(4S)$ resonance separately. 
The yields below the $\Upsilon(4S)$ resonance are $104\pm 14$ for $\rho^0\rho^0$ and $14\pm 4$ 
for $\phi\rho^0$, 
consistent with the expected values of $112\pm 4$ and $13\pm 1$ obtained by scaling the on-peak 
yields of $1138\pm 42$ and $135\pm 13$ by the relative integrated luminosities.

\begin{figure}
\begin{center}
\includegraphics[width=8.8cm]{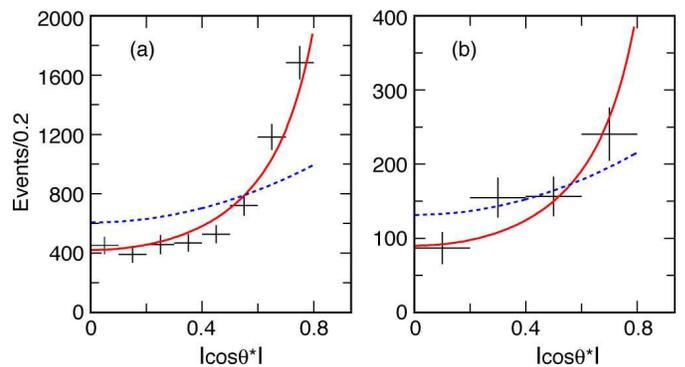}
\caption{\small 
 Production angle distributions, after correction for efficiency, for a) $\rho^0\rho^0$ and b)
 $\phi\rho^0$.
 The solid and dashed lines are the normalized 
$\frac{1+\cos^2\theta^*}{1-\cos^2\theta^*}$ 
and  
$1+\cos^2\theta^*$ distributions, respectively. 
}
\label{fig:myproduction}
\end{center}
\end{figure}

To investigate the production mechanism, we examine the production angle $\theta^*$, 
defined as the angle between  the $\rho^0_f$ ($\phi$)  direction and 
the $e^-$ beam direction in the  CM frame.  
To measure the angular distributions, we subdivide the data into bins of  
$\theta^*$, and repeat the above fit, with linear background slopes $s_{\rho^0_f}$ and $s_{\rho^0_b}$ 
($s_{\rho^0}$ and $s_{\phi}$) fixed to the values from the overall fit. 
The $|\cos\theta^*|$ distributions after MC
efficiency correction are shown in Fig.~\ref{fig:myproduction}. 
The measurements are  restricted to the fiducial region $|\cos\theta^*|<0.8$, 
as the efficiency drops rapidly beyond $0.8$. 
These forward peaking $\cos\theta^*$ distributions are consistent with the 
TVPA expectation which we find can be approximated by: 
\begin{equation}
\label{prodangle}
\frac{d\sigma}{d\cos\theta^*}\propto\frac{1+\cos^2\theta^*}{1-\cos^2\theta^*}
\end{equation}
in the fiducial region.  
The TVPA hypothesis gives a 
$\chi^2/\mathrm{dof}$ (degrees of freedom) of $11.8/7$ ($\rho^0\rho^0$) 
and $3.5/3$ ($\phi\rho^0$).
The fits disfavor  $1+\cos^2\theta^*$,  
giving a $\chi^2/\mathrm{dof}$ of 112/7 for $\rho^0\rho^0$  
and 6.3/3 for $\phi\rho^0$.

\begin{figure}
\begin{center}
\includegraphics[width=8.8cm]{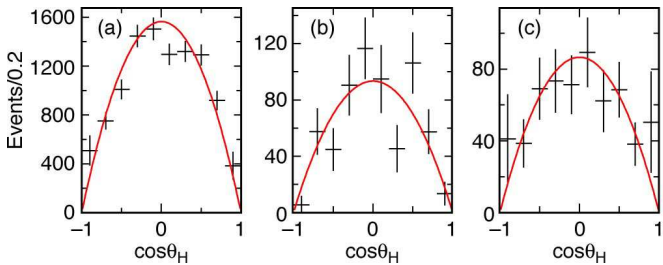}
\caption{\small 
Decay helicity angle distributions for
a) $\rho^0$ from $\rho^0\rho^0$,
b) $\phi$ from $\phi\rho^0$, 
c) $\rho^0$ from $\phi\rho^0$.   
The solid lines are the normalized $\sin^2\theta_H$ distributions.
}
\label{fig:myhelicity}
\end{center}
\end{figure}

Other  observables are the $\phi$ ($\rho^0$) decay helicity angles 
$\theta_H$, defined as the angle, measured in the $\phi$ ($\rho^0$)  
rest frame, between the positively charged kaon or pion and the flight direction 
of the $\phi$ or $\rho^0$ in  the CM frame. 
The efficiency-corrected distribution of $\cos\theta_H$, obtained using the
procedure outline above for $\theta^*$, is shown for the $\rho^0$ and $\phi$
candidates in Fig.~\ref{fig:myhelicity}.
The solid lines in 
Fig.~\ref{fig:myhelicity} are normalized  $\sin^2\theta_H$ distributions   
which give  $\chi^2/\mathrm{dof}$ of $19.3/9$ ($\rho^0$ from $\rho^0\rho^0$),  
$16.4/9$ ($\phi$ from $\phi\rho^0$), 
and  $3.1/9$ ($\rho^0$ from  $\phi\rho^0$).   
The  $\sin^2\theta_H$ distributions indicate that $\phi$ and $\rho^0$ 
are  transversely polarized as expected for TVPA.
The dihedral angles, the angles between the decay planes of the two vector mesons 
measured in the CM frame, 
are consistent with a flat distribution with   $\chi^2/\mathrm{dof}$ of $7.0/9$ ($\rho^0\rho^0$) 
and $10.9/9$ ($\phi\rho^0$). 

The combined hardware and software trigger efficiencies for 
signal events in the fiducial region are 99.9\% for $\rho^0\rho^0$ and 91.3\% for $\phi\rho^0$. 
The lower efficiency for $\phi\rho^0$ is due to an event shape cut in the software trigger.    
For the determination of signal cross sections, the MC  
$\cos\theta^*$ and  $\cos\theta_{H}$ distributions for $\phi$ and $\rho^0$ 
are re-weighted to reproduce the expectation from TVPA.
The signal efficiencies in the fiducial region of $|\cos\theta^*|<$0.8 for 
$\rho^0\rho^0$ and  $\phi\rho^0$ 
are estimated to be 
26.7\% and 23.2\%, 
respectively, including corrections to MC 
simulations of PID, tracking, hardware and software trigger efficiencies.
Initial state photon  radiation is included in the MC simulation.

Systematic uncertainties due to PID and  tracking 
efficiency  are estimated based on measurements
from control data samples. The related  systematic 
uncertainties on lepton vetoes  are estimated by the difference from not 
applying the $e$ and $\mu$ vetoes on pions. 
The  systematic uncertainty from background subtraction  is estimated 
by varying assumptions about background shapes.
We  investigated possible 
feed-down background from related  modes with an extra $\pi^0$ using
various extrapolations from the four-particle mass sidebands. 
We assume that the final states are fully transversely polarized.
The systematic uncertainties are summarized in Table ~\ref{tab:syserror}.

Taking the branching fraction of $\phi\rightarrow$$K^+K^-$ as 49.1\% and 
$\rho^0\rightarrow$$\pi^+\pi^-$ as 100\% ~\cite{pdg04}, and signal mass regions of 
$0.5< m_{\rho^0}<1.1$\gevcc and $1.008< m_{\phi}<1.035$\gevcc, 
we obtain the following results for the TVPA cross sections within 
$|\cos\theta^*|<$0.8 near $\sqrt{s}=10.58\gev$: 
\begin{eqnarray*} 
 \sigma_{\mathrm{fid}}(e^+e^-\rightarrow\rho^0 \rho^0) & = & \Resultrhozerorhozeroformal \\
 \sigma_{\mathrm{fid}}(e^+e^-\rightarrow\phi \rho^0) & = & \Resultphirhozeroformal .
\end{eqnarray*}
The measured cross sections are in good agreement with  the calculation from a vector-dominance 
two-photon exchange model~\cite{davier}.

\par In summary, we have observed  exclusive production of $C=+1$ final 
states in $e^+e^-$ interactions. 
The measured $C$ parity configuration, the signal yields 
in data samples on the $\Upsilon(4S)$ resonance and below, and the production angle distributions 
support the conclusion that the production mechanism is 
two-virtual-photon annihilation. 
The Standard Model predictions of the anomalous magnetic moment of the muon  and 
the QED coupling  rely on 
the measurements of low-energy $e^+e^-$ hadronic cross sections, 
which are assumed to be entirely due to single-photon exchange. 
We have estimated the effect  due to the TVPA processes 
we have measured, and find it to be small compared with the current precision~\cite{g-2}.

\begin{table}
\begin{center}
\caption{Systematic uncertainties on the cross sections for $\epem \rightarrow \rho^0\rho^0$/$\phi\rho^0$.}
\begin{tabular}{lrr}
\hline\hline
                        & $\rho^0\rho^0$ & $\phi\rho^0$ \\
\hline
Particle Identification                & 9.6\%         &  10.4\%     \\	
 Background subtraction                & 7.0\%        &  7.0\%     \\	
Tracking efficiency                    & 5.0\%         &  5.0\%     \\	
 $\rho^0\rho^0\pi^0,\phi\rho^0\pi^0$ background & 1.6\%         &  2.7\%     \\
 Luminosity                            & 1.2\%         &  1.2\%     \\
\hline 
 Total                                 & 13.0\%        &  14.0\%     \\                
\hline\hline
\end{tabular}    
\label{tab:syserror}
\end{center} 
\end{table}

\par We are grateful for the excellent luminosity and machine 
conditions
provided by our \pep2\ colleagues, 
and for the substantial dedicated effort from
the computing organizations that support \babar.
We wish to thank S. Brodsky, M. Peskin and H. Quinn for helpful 
discussions. The collaborating institutions wish to thank 
SLAC for its support and kind hospitality. 
This work is supported by
DOE
and NSF (USA),
NSERC (Canada),
IHEP (China),
CEA and
CNRS-IN2P3
(France),
BMBF and DFG
(Germany),
INFN (Italy),
FOM (The Netherlands),
NFR (Norway),
MIST (Russia), and
PPARC (United Kingdom). 
Individuals have received support from CONACyT (Mexico), A.~P.~Sloan Foundation, 
Research Corporation,
and Alexander von Humboldt Foundation.


\begin{thebibliography}{99}


\bibitem{g-2}
M. Davier, S. Eidelman, A. Hoecker, Z. Zhang, \epjc{31}, 503 (2003).


\bibitem{babardetector}
\babar~ Collaboration, B. Aubert {\em et al.}, Nucl. Instrum. Methods, Sect. A {\bf 479}, 1 (2002).


\bibitem{pdg04}
Particle Data Group, S.~Eidelman {\em et al.}, \plb{592}, 1 (2004).


\bibitem{MC}
$\epem\rightarrow \rho^0\rho^0/\phi\rho^0 $ are generated uniformly over phase space.  
We use the AFKQED package to simulate the signal processes, including hard and multiple soft
initial state radiation, using the methods in~\cite{ckhhad}
and~\cite{strfun}.


\bibitem{ckhhad}
H.~Czy\.z and J.H.~K\"uhn, Eur.\ Phys.\ J.\ C {\bf 18}, 497 (2001).
                                                                     
\bibitem{strfun}
M.~Caffo, H.~Czy\.z, and E.~Remiddi, Nuo.\ Cim.\ {\bf 110A}, 515
(1997);
Phys.\ Lett.\ B {\bf 327}, 369 (1994).

\bibitem{davier}
M.~Davier, M.~Peskin, and A.~Snyder, hep-ph/0606155.



\end{thebibliography}
\end{document}